\documentclass{ifacconf}

\usepackage{graphicx}      
\usepackage{natbib}        
\usepackage{amsmath,amsfonts,amssymb}
\usepackage{tikz}
\usepackage{xcolor}
\usepackage{soul}

\begin{document}
\begin{frontmatter}

\title{A simulation-based comparative analysis of PID and LQG control for closed-loop anesthesia delivery} 

\author[A,First,Second]{Sourish Chakravarty},
\author[A,First,Second]{Ayan S. Waite},
\author[First,Second]{John H. Abel},
\author[First,Second,Third]{Emery N. Brown}

\address[First]{Picower Institute for Learning and Memory, Massachusetts Institute of Technology (MIT), Cambridge, MA}
\address[Second]{Department of Anesthesia, Critical Care and Pain Medicine at Massachusetts General Hospital, Boston, MA}
\address[Third]{Institute for Medical Engineering and Science, MIT, Cambridge, MA}
\thanks[A]{S.C. and A.S.W. contributed equally.}
\thanks[footnoteinfo]{This work was partially supported by NIH Award P01-GM118629 (to E.N.B.), by funds from Massachusetts General Hospital (to E.N.B.), by funds from the Picower Institute for Learning and Memory (to E.N.B. and S.C.),  and by NIH/NIA Award F32-AG064886 (to J.H.A.). This work has been accepted by IFAC for publication \textcopyright IFAC 2020.}

\begin{abstract}                
Closed loop anesthesia delivery (CLAD) systems can help anesthesiologists efficiently achieve and maintain desired anesthetic depth over an extended period of time. A typical CLAD system would use an anesthetic marker, calculated from physiological signals, as real-time feedback to adjust anesthetic dosage towards achieving a desired set-point of the marker.  Since control strategies for CLAD vary across the systems reported in recent literature, a comparative analysis of common control strategies can be useful. For a nonlinear plant model based on well-established models of compartmental pharmacokinetics and sigmoid-Emax pharmacodynamics, we numerically analyze the set-point tracking performance of three output-feedback linear control strategies:  proportional-integral-derivative (PID) control, linear quadratic Gaussian (LQG) control, and an LQG with integral action (ILQG). Specifically, we numerically simulate multiple CLAD sessions for the scenario where the plant model parameters are unavailable for a patient and the controller is designed based on a nominal model and controller gains are held constant throughout a session. Based on the numerical analyses performed here, conditioned on our choice of model and controllers, we infer that in terms of accuracy and bias PID control performs better than ILQG which in turn performs better than LQG. In the case of noisy observations, ILQG can be tuned to provide a smoother infusion rate while achieving comparable steady state response with respect to PID. The numerical analysis framework and findings reported here can help CLAD developers in their choice of control strategies. This paper may also serve as a tutorial paper for teaching control theory for CLAD.
\end{abstract}

\end{frontmatter}
\section{Introduction}
More than 230 million patients undergo major surgeries world-wide under anesthesia (\cite{weiser2008estimation}). During major surgeries performed under general anesthesia (GA) (\cite{brown2010general}), conventionally, an anesthesiologist would monitor relevant physiological signals that are known to be associated with anesthetic depth, and manually adjust the drug dose. For example, changes in oscillatory patterns in the scalp electroencephalogram (EEG) has been shown to be correlated with behaviorally-defined changes in unconsciousness levels during GA (\cite{purdon2013electroencephalogram}). Such structured drug-dependent changes in the EEG signals have allowed developments of automated EEG-based closed-loop anesthesia delivery (CLAD) systems that can provide precise control of brain states under anesthesia (\cite{bickford1950automatic, schwilden1989closed,absalom2002closed,dumont2009robust, shanechi2013brain,puri2016multicenter,yang2019developing}). Analogous systems are also available for blood pressure (BP)-based CLAD (\cite{ngan2007closed}). 

A typical CLAD system works in a cyclic mode at a fast rate ($\sim 0.1 \, s^{-1}$) where in each cycle a computer calculates a scalar anesthetic marker from real-time physiological signals (say, EEG or BP) and uses a feedback control strategy to determine the adjustment of drug-dosage for the next cycle. As this iterative procedure continues over time, the anesthetic marker ideally would approach the user-prescribed set-point. The utility of such {\it autopilot} systems are in personalized medicine as they can aid anesthesiologists achieve patient-specific precise drug-dosing efficiently, even under resource constrained environments (\cite{dumont2013closed, absalom2011closed}). 

The CLAD systems reported in prior research, vary with respect to different features, e.g., regime of anesthesia, anesthetic drug, signal modality, marker definition, control strategy, and plant model, just to name a few. For example, the classical output-feedback proportional-integral-derivative (PID) control strategy and its variants have been found to be useful in CLAD applications in human studies (\cite{dumont2009robust, puri2016multicenter,westover2015robust}). The more recent development of using an estimated-state feedback-based optimal control strategy via linear quadratic Gaussian (LQG) regulation have yielded promising results for maintenance of medically-induced coma in animal models (\cite{shanechi2013brain}). The success of both PID and LQG, as general control paradigms for CLAD, motivated us to compare them through a numerical experiment in this work.

In this numerical study of CLAD, we consider a specific nonlinear state-space model as the plant model. Here the state dynamics, driven by the drug infusion, is described by a four-compartment linear pharmacokinetic (PK) model. The observation dynamics, relating the anesthetic marker and the dynamic state in the effect compartment, is assumed to be described by a {\it sigmoid-Emax} pharmacodynamic (PD) model (\cite{meibohm1997basic}). We choose realistic values of model parameters based on a pair of well-established models in literature for propofol-based PK and corresponding EEG-based PD (\cite{schnider1998influence, schnider1999influence}). We linearize this nonlinear plant model around a steady state value and use the resulting linear state-space model to determine control parameters. Since LQG is known to have robustness issues (\cite{doyle1978guaranteed}) and model-misspecification can lead to steady state errors during set-point tracking, we also consider LQG with an additional integral action (ILQG) on the output error. To address practical constraints, we impose actuator saturation and zero-order holds explicitly. We consider different operating conditions characterized by observation noise and pump-update rates, as well as control parameters. For fixed target trajectory and a given operating condition, we simulate CLAD for multiple subjects to mimic a practical scenario when only a nominal plant model, instead of the true one, is available for controller design. 

In the ensuing discussion, Sec.~\ref{sec:methods} presents the core equations for the plant and the control strategies. Sec.~\ref{sec:results} presents our numerical simulation design, results and associated discussion. Finally, Sec.~\ref{sec:conclusion} summarizes our work. 
\vspace{-5pt}
\section{Methods}\label{sec:methods} 
\vspace{-5pt}\subsection{Nonlinear plant model}
\begin{figure}[htbp]
    \centering
    \includegraphics[width=0.4\textwidth]{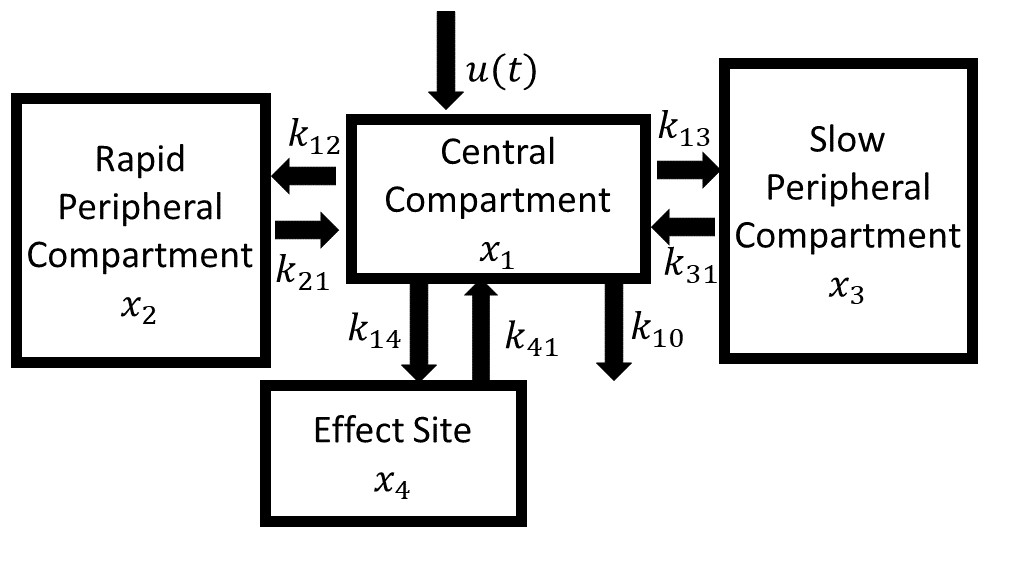}
        \vspace{-20pt}
    \caption{Schematic of a 4-compartment PK model}
    \label{fig:schema_PKPD}
\end{figure}

We model the pharmacokinetics (PK) using a first-order ordinary differential equation as, 
\begin{equation}
    \dot{x}(t) = Ax(t) + Bu(t)
    \label{eq:LinearPK}
\end{equation}
where, $\dot{x}$ indicates $dx/dt$. Transition matrix, $A$, and input scaling matrix, $B$, are prescribed based on a four compartment mammillary model assumption as proposed by \cite{shafer1992algorithms}, where the effect site is augmented to a three-compartment model as an additional fourth compartment (much smaller in volume relative to the central compartment). The mass of drug in all the compartments at any instant of time, $t$, is indicated by $x(t) =[x_1(t), x_2(t), x_3(t), x_4(t)]^T$, $u(t)$ denotes scalar infusion rate function, $B=[1, 0, 0, 0]^T$ and transition matrix is given by,
\begin{align}
            A
            = 
	\left[
	\begin{matrix}
                - (k_{10} + k_{12} + k_{13}+k_{14}) & k_{21} & k_{31} & k_{41}\\
                k_{12} & -k_{21} & 0 & 0\\
                k_{13} & 0 & -k_{31} & 0 \\
                k_{14} & 0 & 0 & -k_{41}
	\end{matrix}
	\right]
    \label{eq:Kmat}
\end{align}
Note, that $u(t)\ge0$ and $x_i(t)\ge0$ $\forall\, i\in \lbrace 1,2,3,4\rbrace$ and $\forall\, t\ge 0$. The PK model described above is illustrated in Fig.~\ref{fig:schema_PKPD}. We model the pharmacodynamic (PD)\footnote{Conventionally, $k_{14}$ and $k_{41}$ are treated as PD parameters (\cite{schnider1999influence}), but here, for convenience, we consider them as PK model parameters so as to maintain separate parameter sets for the state and the observation equations.} relationship between a scalar-valued dose-dependent function, say $y(t)$, of the physiological signal of interest and the effect site amount, $x_4(t)$, using the following sigmoid-Emax model,
\begin{align}
    y(t) = E_0 + E_{max} \left( 
    \frac{ (x_4(t)/x_{50})^\gamma }{1 + (x_4(t)/x_{50})^\gamma}
    \right)
    \label{eq:sigmoidEmaxPD}
\end{align}
where, $E_0$, $E_{max}>0$, $x_{50}>0$, and $\gamma>0$ are scalar parameters.  In the ensuing discussion, the symbol `$(t)$', in the notations indicating dynamic variables, will be omitted for brevity. 
In the computational simulation studies that we perform here, the core plant model taking input $u(t)$ and generating output $y(t)$ is given by Eqs.~\eqref{eq:LinearPK} and \eqref{eq:sigmoidEmaxPD}. The nonlinear plant model is characterized by the subject-specific PK-PD parameters $\Theta_P = \lbrace A, x_{50}, \gamma, E_{max}, E_0 \rbrace$.
\subsection{\label{subsec:LinControl}Linear control strategy with output feedback}
\vspace{-5pt}
\subsubsection{\label{sec:SteadyState}Steady state relationships:}
For linear control design for set-point tracking, we determine relationships among $y$, $x$ and $u$ for a steady state condition. Imposing the steady state condition, $\dot{x}=0$ in Eq.~\eqref{eq:LinearPK}, and denoting the constant values at steady state as $y_{ss}$, $x_{ss}$ and $u_{ss}$, we obtain
\begin{align}
    & x_{ss} = -A^{-1}Bu_{ss} \,,
    \label{eq:xss_uss}
    \\
    & x_{4,ss} = [0, 0, 0, 1]x_{ss}\,.
\end{align}
$y_{ss}$ can be calculated by substituting $x_{4,ss}$ in Eq.~\eqref{eq:sigmoidEmaxPD}. When either $y_{ss}$ or $u_{ss}$ is specified, one can use the aforementioned algebraic relationships to calculate the rest of the steady state quantities.
\vspace{-5pt}
\subsubsection{Linearizing about a steady state:}
The relationship between the $y$ and $u$, via Eqs.~\eqref{eq:LinearPK} and \eqref{eq:sigmoidEmaxPD}, is nonlinear. To incorporate a linear feedback control framework to regulate $y$ around a steady state value, characterized by the tuple $(y_{ss}, x_{ss}, u_{ss})$, we linearize  Eq.~\eqref{eq:sigmoidEmaxPD} about this steady state,
\begin{align}
    y-y_{ss} = \left(\frac{dy}{dx_4}\right)_{x_{4,ss}}
    (x_4 - x_{4,ss}) + \text{higher order terms}
\end{align}
Ignoring the higher-order terms, we obtain a new set of equations in terms of dynamic quantities that denote perturbations with respect to respective steady state values,
\begin{align}
&    \dot{\tilde{x}} = A\tilde{x} + B\tilde{u}
\label{eq:linearPK_perturb}
\\
&
\tilde{y} = C\tilde{x} 
\label{eq:linearPD_perturb}
\end{align}
where, $\tilde{y}\equiv y - y_{ss}$, $\tilde{x}\equiv x - x_{ss}$, $\tilde{u}\equiv u - u_{ss}$, and 
\begin{align}
    C = 
    \begin{bmatrix}
    0 & 0 &
    0 & \left(\frac{dy}{dx_4}\right)_{x_{4,ss}}
    \end{bmatrix}
\end{align} 
\\
\begin{align}
\text{where, }
    \left(\frac{dy}{dx_4}\right)_{x_{4,ss}} = \frac{ E_{max}}{x_{50}}
    \frac{ \gamma  (x_{4,ss}/x_{50})^{\gamma-1} }{(1 + (x_{4,ss}/x_{50})^\gamma)^2}
    \label{eq:C_mat}
\end{align}
\vspace{-5pt}
\subsubsection{PID control:}
To achieve set-point tracking of $y$ about a constant set-point $y_{sp}$, we consider linear system of equations in Eqs.~\eqref{eq:linearPK_perturb} and \eqref{eq:linearPD_perturb} where linearization is achieved around a user-prescribed state $(y_o,\,x_o,\,u_o)$. Using PID control gains, $K_P$, $K_I$ and $K_D$, tuned based on this linear system, we implement the following control strategy (\cite{astrom2010feedback}),
\begin{align}
    u_{pid} = u_{o} + K_P (y-y_{sp}) + K_I \int (y-y_{sp}) dt + K_D \dot{y}
\end{align}
where, $y_{sp}$ denotes the user-prescribed set-point which may or may not be equal to $y_{o}$, but should be in the neighborhood around $y_{o}$ where the linear dynamics regime can still be assumed. The parameter set for the PID is given by $\Theta_{PID}=\lbrace K_{P}, K_{I}, K_{D}, A, x_{50}, \gamma, E_{max}, E_{0}, u_{o} \rbrace$. This parameter set includes controller gains ($ K_{P}, K_{I}, K_{D}$) as well as the parameters of the nominal linear model that will depend both on the nonlinear plant parameters ($A, x_{50}, \gamma, E_{max}, E_{0}$) as well as the state ($u_{o}$) at which the linearization occurs.
\vspace{-5pt}
\subsubsection{LQG control:}
To achieve constant set-point tracking when using LQG strategy, we reformulate the problem in the form of,
\begin{align}
    & \dot{x} = A(x-x_{sp}) + B(u-u_{sp}) + w
    \label{eq:LQG_state}
    \\
    &
    y - y_{sp} = C_{o} (x-x_{sp}) + v
    \label{eq:LQG_obs}
\end{align}
where, $w$ and $v$ are continuous-time zero-mean Gaussian white noise processes characterized by positive-definite covariance matrix $W$ and positive scalar $V$, respectively (\cite{crassidis2011optimal}). The $C_o$ is determined from Eq.~\eqref{eq:C_mat} using the linearization state $(y_o, x_o, u_o)$ prescribed in terms of either $y_o$ or $u_o$ (see Sec.~\ref{sec:SteadyState}). We assume that during a session the controller parameters that depend on $(y_o, x_o, u_o)$ will be kept constant, but the set-point $y_{sp}$ may vary around $y_o$ in a session. With the problem setup in Eqs.~\eqref{eq:LQG_state} and \eqref{eq:LQG_obs}, we can pose the LQG optimal control problem as,
\begin{align}
    \lbrace u^\star(t) \rbrace = \mathrm{argmin}\, E\left[ J(\lbrace u(t) \rbrace) \right]
\end{align}
\begin{align}
     J \equiv \int_{0}^{\infty}
    (x-x_{sp})^T Q(x-x_{sp}) + (u-u_{sp})^T R (u-u_{sp}) \, dt
    \label{eq:LQG_Obj}
\end{align}
such that Eq.~\eqref{eq:LQG_state} holds at every time-point, and $E[\cdot]$ indicates an expectation operation. Assuming the system $(A, B)$ is stabilizable and $(A, C_o)$ is detectable and matrices $Q$ and $R$ are positive semi-definite and positive definite, respectively, the solution of the aforementioned optimal control problem leads to a steady state linear quadratic regulator with control action $ u_{lqg} $ given by,
\begin{align}
    u_{lqg} = u_{sp} -K_C(\widehat{x} - x_{sp})
    \label{eq:LQG_u}
\end{align}
where, $\widehat{x}$ is the output of a steady state Kalman filter,
\begin{align}
    \dot{\widehat{x}} = (A-K_E C_o)(\widehat{x}-x_{sp}) + B(u-u_{sp)} + K_E (y-y_{sp})
    \label{eq:LQG_xhat}
\end{align}
Controller gain matrix $K_C$ depends on $A$, $B$, $Q$ and $R$. Estimator gain matrix $K_E$ depends on $A$, $C_{o}$, $W$ and $V$. The formulae to calculate $K_C$ and $K_E$ can be found in literature (\cite{athans1971role, crassidis2011optimal}). The parameter set for the PID is given by $\Theta_{LQG}=\lbrace Q, R, W, V, A, x_{50}, \gamma, E_{max}, E_{0}, u_{o} \rbrace$.
\vspace{-5pt}
\subsubsection{ILQG control:}
To minimize any steady state bias that can ensue due to model mismatch between the plant and the controller, we add an integral action to minimize the steady state errors. To do this we consider an additional state $z$ driven by the tracking error,
\begin{align}
    & \dot{z} =  y_{sp} - y\,,\text{ such that},
    \\
    & u_{ilqg} = u_{lqg} + K_I  z 
\end{align}
where, $K_I>0$ is an additional user-prescribed gain. The parameter set for the ILQG is given by $\Theta_{ILQG}=\lbrace K_I, Q, R, W, V,$ $A, x_{50}, \gamma, E_{max}, E_{0}, u_{o} \rbrace$. Since we are interested in how the LQG performance changes with regard to the current CLAD problem in the presence and absence of an integral action, we explicitly specify $K_I$. This is in the same spirit as the LQG servocontroller designs that can be simulated using the \textit{lqgtrack} function in \cite{MATLAB:2019b}. Alternatively, one can also design $K_I$ implicitly within the LQG framework (\cite{grimble1979design}). 
\begin{figure}
    \centering
    \includegraphics[width=0.475\textwidth]{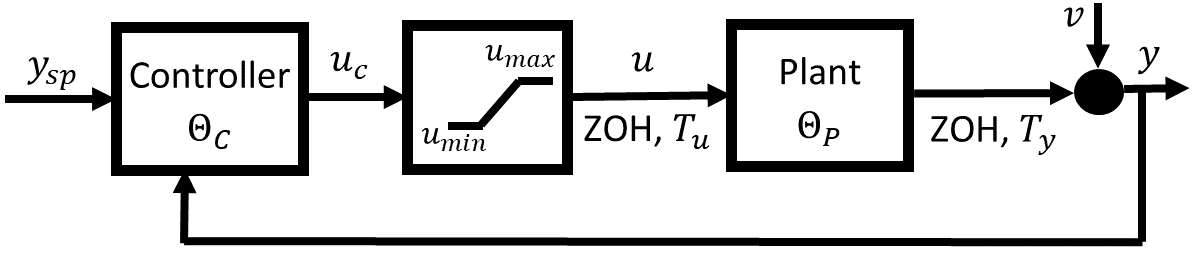}
    \vspace{-5pt}
    \caption{Schematic block diagram of CLAD simulation setup}
    \label{fig:CTRL_BLOCK}
\end{figure}
\vspace{-5pt}
\subsubsection{{Additional constraints for CLAD simulation}:}
In the context of CLAD, practical considerations require $u\ge u_{min}=0 $ and $u\le u_{max}$; the latter constraint can be due to constraints on the physiological system and/or the mechanical actuation system. To mimic this in our simulation, we allow output of the controller, say $u_c$ (such that $c\in \lbrace pid, lqg, ilqg\rbrace$), goes through a saturation block prescribed as,
\begin{align}
    & u = u_c \,\, , \text{ if } 0<u_c<u_{max},
    \nonumber
\\
    & u = 0 \,\, , \text{ if } u_c\le 0,
    \nonumber
\\
    & u = u_{\max} \,\,, \text{ if } u_c\ge u_{\max}, 
\end{align}
to yield the control signal $u$ that is input to the drug delivery system. Within the drug delivery system, the modulation of infusion rates for anesthetic fluids is a mechanical phenomenon which can have a short transient before the infusion rate stabilizes around the desired flow rate. To simulate this condition without modeling the drug delivery dynamics explicitly we impose a \textit{zero-order hold} (zoh) condition on $u$, such that the flow rate $u$ is updated at an interval $T_{u}$, and held constant until the next update. A similar zoh condition is applied on the output side assuming that it is sampled at rate $T_y$ which may or may not be equal to $T_u$. To mimic the noise in the observations, we also add Gaussian white noise with variance $V$ and sampled at every interval $T_y$, to the observation. If $T_y = T_u$ holds, then one can alternatively choose to model plant and controller dynamics in discrete-time governed by incremental dynamic equations. However, this would require the increment time-interval to enter the system matrices that are used to calculate controller gains. Furthermore, modeling the plant in continuous time is conceptually closer to the physiological phenomena governing the true PK-PD phenomena. Therefore, in this work, we try to keep the plant and control blocks independent of any prescribed time-interval, but add the aforementioned constraints, via user-prescribed values of $u_{max}$, $T_u$ and $T_y$, externally in the complete simulation block diagram (Fig.~\ref{fig:CTRL_BLOCK}).
\begin{table}[htbp]
    \vspace{-5pt}
    \centering
    \caption{
     {Performance metrics for PID} (Fig.\ref{fig:PID_simu}(a-d)), LQG with $\rho=10^{-13}$ (Fig.\ref{fig:LQG_simu}(a-d)), LQG with $\rho=10^{-6}$ (Fig.\ref{fig:LQG_simu}(e-h)), ILQG with $K_I = 10, \rho=10^{-13}$ (Fig.\ref{fig:LQG_IA_simu}(a-d)) , LQG with $K_I = 1, \rho=10^{-6}$ (Fig.\ref{fig:LQG_IA_simu}(e-h)). Median (min, max) values of performance metrics from 44 CLAD simulations per case.
   } 
    \begin{tabular}{p{0.0175\textwidth}|p{0.025\textwidth}|p{0.025\textwidth}|p{0.09\textwidth}|p{0.110\textwidth}|p{0.075\textwidth}}
    \hline
  Fig. &  $60T_u$ & $V$  & $\epsilon_I\,(\%)$ &  $\epsilon_B\,(\%)$ &  $u_{avg}$ \\
\hline
\ref{fig:PID_simu}(a)& 5s& $10^{-6}$&
0.4(0.4,0.4)
&
0.1(0.1,0.1)
&
5.6(4.0,7.3)\\
\ref{fig:PID_simu}(b)& 5s & $10^{-4}$ &
3.2(3.2,3.3)
&
0.2(0.1,0.2)
&
5.6(4.0,7.3)\\
\ref{fig:PID_simu}(c)& 10s& $10^{-6}$&
0.3(0.3,0.4)
&
0.1(0.1,0.1)
&
5.7(4.1,7.3)\\
\ref{fig:PID_simu}(d)& 10s & $10^{-4}$&
2.4(2.3,2.4)
&
0.2(0.2,0.3)
&
5.6(4.0,7.3)\\   
    \hline
\ref{fig:LQG_simu}(a)&5s& $10^{-6}$  & 
2.6(1.2,6.7) 
& 
-2.5(-6.6,1.6) 
& 
5.4(4.2,6.6)\\
\ref{fig:LQG_simu}(b)&5s&  $10^{-4}$ & 
4.2(3.7,7.0) 
& 
-2.7(-6.8,1.7) 
& 
5.4(4.2,6.6)\\
\ref{fig:LQG_simu}(c)&10s& $10^{-6}$ &
2.6(1.5,6.5) 
&
-2.4(-6.4,1.4) 
&
5.4(4.2,6.6)\\
\ref{fig:LQG_simu}(d)&10s& $10^{-4}$ &
3.6(3.0,6.8) 
&
-2.6(-6.6,1.9) 
&
5.4(4.2,6.6)
\\
\hline
\ref{fig:LQG_simu}(e)&5s& $10^{-6}$ &
16.9(3.4,47.2)
&
-16.9(-47.2,4.7)
&
4.6(4.6,4.6)\\
\ref{fig:LQG_simu}(f)&5s& $10^{-4}$ &
14.8(5.5,42.8)
&
-14.8(-42.8,7.0)
&
4.6(4.6,4.6)\\
\ref{fig:LQG_simu}(g)&10s& $10^{-6}$ &
17.2(3.4,47.5)
&
-17.0(-47.5,4.3)
&
4.6(4.6,4.6)\\
\ref{fig:LQG_simu}(h)&10s& $10^{-4}$ &
15.2(4.7,44.7)
&
-15.2(-44.7,7.0)
&
4.6(4.6,4.6)\\
\hline
\ref{fig:LQG_IA_simu}(a)&5s& $10^{-6}$& 
1.0(0.9,1.2)
&
0.5(0.2,0.9)
&
5.6(4.1,7.2)\\
\ref{fig:LQG_IA_simu}(b)&5s& $10^{-4}$& 
3.6(3.5,3.7)
&
0.3(-0.3,1.2)
&
5.6(4.1,7.2)\\
\ref{fig:LQG_IA_simu}(c)&10s& $10^{-6}$&
1.0(1.0,1.2)
&
0.5(0.1,1.0)
&
5.7(4.1,7.2)\\
\ref{fig:LQG_IA_simu}(d)&10s& $10^{-4}$
&
2.9	(2.8,3.0)
&
0.5(-0.1,1.3)
&
5.7(4.1,7.2)\\
\hline
\ref{fig:LQG_IA_simu}(e)&5s& $10^{-6}$&
1.2(0.9,1.6)
&
0.7(0.5,0.9)
&
5.6(4.1,7.3)\\
\ref{fig:LQG_IA_simu}(f)&5s& $10^{-4}$&
3.7(3.5,3.9)
&
0.6(0.3,0.9)
&
5.6(4.1,7.3)\\
\ref{fig:LQG_IA_simu}(g)&10s& $10^{-6}$&
1.5(1.0,1.9)
&
0.8(0.6,1.0)
&
5.7(4.1,7.3)\\
\ref{fig:LQG_IA_simu}(h)&10s& $10^{-4}$&
2.8(2.6,3.1)
&
0.8(0.5,1.1)
&
5.7(4.1,7.3)\\
\hline\hline
    \end{tabular}
    \label{tab:Scenario_B}
\end{table}
\vspace{-5pt}
\section{Results}\label{sec:results}
\vspace{-5pt}
\subsection{CLAD simulation setup}
\vspace{-5pt}
\subsubsection{Plant model parameters:}
As candidate plant model parameters we used the weight, height, age and sex information available for $N_p = 44$ patients reported in prior work \cite[Table A1]{dumont2009robust}.  For the PK-PD model, we used the model parameters reported for 3-compartment propofol pharmacokinetics \cite[Table 2]{schnider1998influence} and corresponding EEG-based pharmacodynamics \cite[Table 3]{schnider1999influence}, where for the latter we use only the age-adjusted parameters for the ascending sigmoidal function in the bi-phasic response \cite[Fig. 1]{schnider1999influence}. We set $E_0 = 0$ and $E_{max}=1$ since they reflect a constant shift and scaling.  Per suggestion by \cite{shafer1992algorithms}, we use $k_{41} = 10^{-5}k_{14}$. 
\begin{figure}[htbp]
    \centering
    \includegraphics[width=0.475\textwidth]{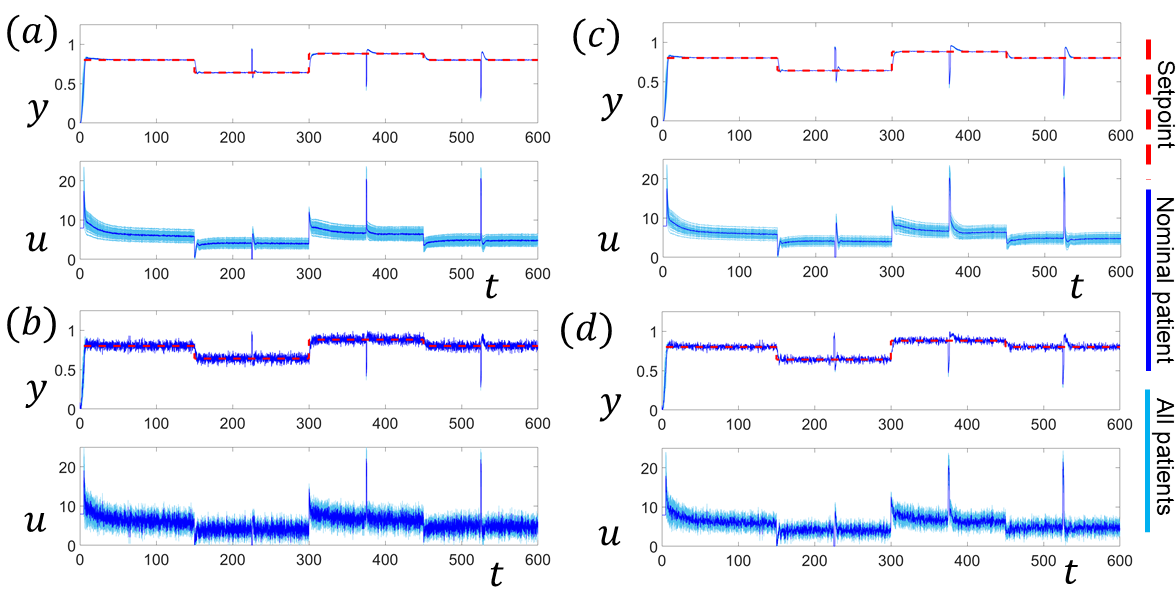}
 \caption{    \label{fig:PID_simu} {\bf PID}: Simulated trajectories for 45 patients (including the nominal subject, highlighted in solid blue) with 4 sets of ($60T_u, V$): (a) ($5\,s, 10^{-6}$), (b) ($5\,s, 10^{-4}$), (c) ($10\,s, 10^{-6}$), (d) ($10\,s, 10^{-4}$). Also see Table~\ref{tab:Scenario_B}.}.
\end{figure}
\vspace{-15pt}
\subsubsection{Simulation parameters:}
Each simulation mimics a CLAD session of duration, $T_{max}=600 \, min$. In all simulations, we maintain the maximum allowable infusion rate for a patient weighing $W\, (kg)$ as $u_{max} = 1 \, mg/kg/min \times W $. We set $T_y = T_u$ for simplicity.
To mimic a practical situation where the CLAD system is activated after a patient is anesthetized, we run the system in an open-loop mode for $T_{ind}=5\, min$ with a fixed infusion rate of $u=u_{max}/10$ prior to activating the closed-loop mode.
For the closed-loop mode, the following target profile is maintained: $y_{sp}=0.8$ for $t\in (T_{ind}, T_{max}/4]$, then $y_{sp}=0.64$ for $t\in (T_{max}(1/4), T_{max}(2/4)]$, followed by $y_{sp}=0.88$ for $t\in (T_{max}(2/4), T_{max}(3/4)]$, and finally back to $y_{sp}=0.9$ for $t\in (T_{max}(3/4), T_{max}]$. We also simulate disturbance profiles at $t=225 \,min$, $375\, min$ and $575\, min$ where the effect-site dynamic state $x_4$ is scaled by a factor 2, 0.5 and 0.5, respectively, for a brief period ($=10T_u$).
\vspace{-5pt}
\subsubsection{Controller parameters:} For any given simulation run we linearize the PK-PD model, available to the controller, around a steady state value of $u_o=u_{max}/10$. For LQG and ILQG, we set $Q$ to be a diagonal matrix such that $ diag(Q) = \left[{1}/{(x_{1,max} - x_{o})^2}, {1}/{(x_{2,max} - x_{o})^2} ,
\right.$ $\left. {1}/{(x_{3,max} - x_{o})^2}, {\rho}/{(x_{4,max} - x_{o})^2}\right]$, where $x_{max}$ is calculated using Eq.~\eqref{eq:xss_uss} and $u_{max}$. We set scalar $R=1/(u_{max}-u_o)^2$. For the process noise variance $W$ we choose a diagonal matrix such that $diag(W) = 10^{-6}V[1, 1, 1, 1]$. For the PID controller, we determine $K_P$, $K_I$ and $K_D$ for linear system parameters $A$, $B$ and $C_o$ by invoking the Matlab function, \textit{pidtune} with its default setting of balance between performance and robustness. Since we will be simulating with noisy observations, we incorporate a derivative filter such that transfer function contribution to the derivative component of the PID transfer function is $K_D s/(T_f s + 1)$ where we set $T_f=5T_u$. Furthermore, since we have actuation saturation, to avoid integral error build-up, we employ the following anti-windup strategy (\cite{astrom2010feedback}): when the output of the controller, $u_c$, falls outside the range $[u_{min}, u_{max}]$ we enforce a zero input into the integrator blocks of both the PID and the ILQG.
\begin{figure}[htbp]
    \centering
    \includegraphics[width=0.475\textwidth]{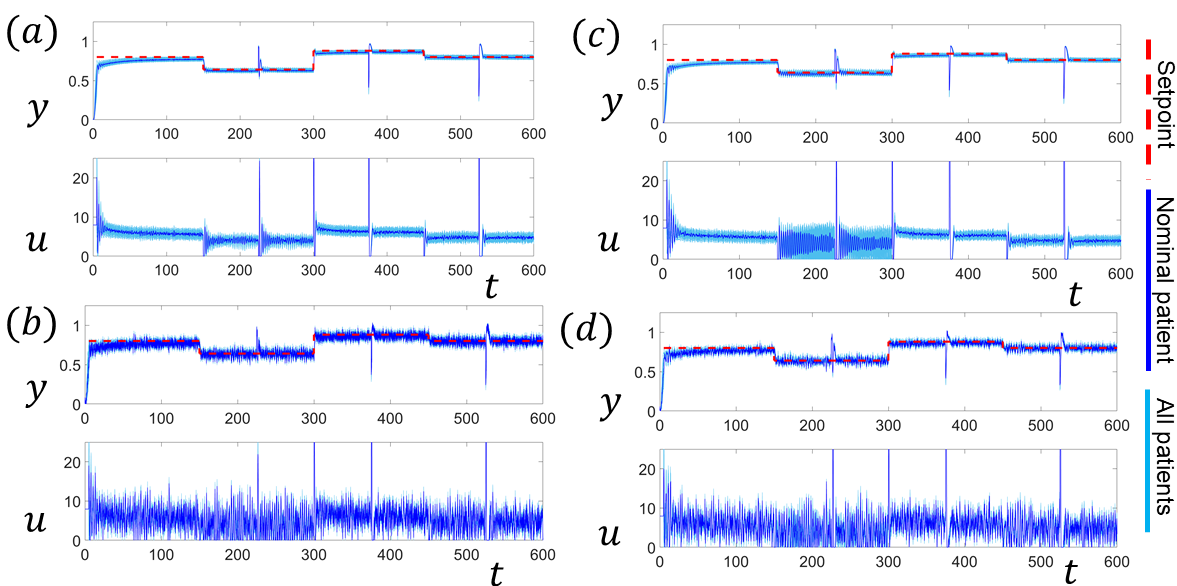}
    \includegraphics[width=0.475\textwidth]{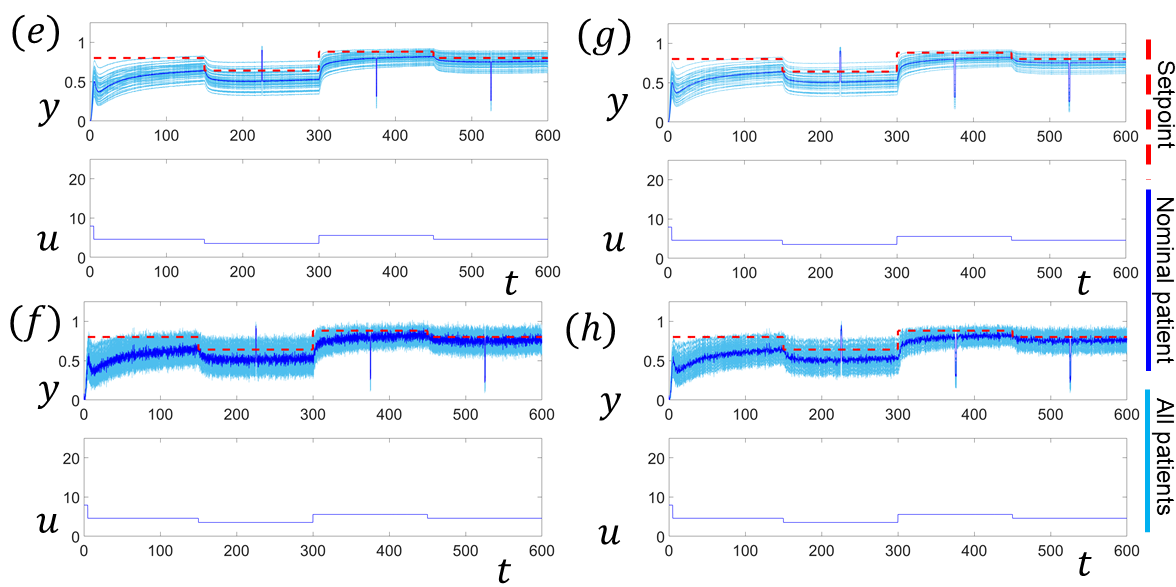}
  \caption{    \label{fig:LQG_simu} {\bf LQG}: Simulated trajectories for 45 patients (including the nominal subject, highlighted in solid blue) with 4 sets of ($60T_u, V$): (a,e) ($5\,s, 10^{-6}$), (b,f) ($5\,s, 10^{-4}$), (c,g) ($10\,s, 10^{-6}$), (d,h) ($10\,s, 10^{-4}$). (a-d) $\rho=10^{13}, K_I = 0$, (e-h) $\rho=10^{6}, K_I = 0$. See Table~\ref{tab:Scenario_B}.}
 \end{figure}
\vspace{-5pt}
\subsection{CLAD simulations}
\begin{figure}[htbp]
    \centering
    \includegraphics[width=0.475\textwidth]{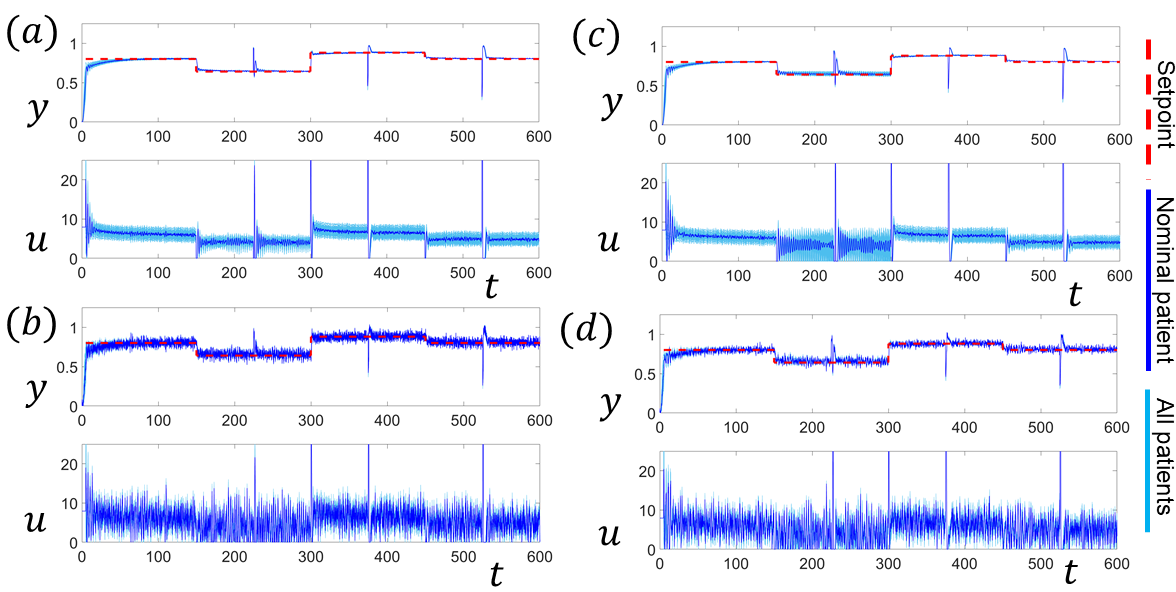}
    \includegraphics[width=0.475\textwidth]{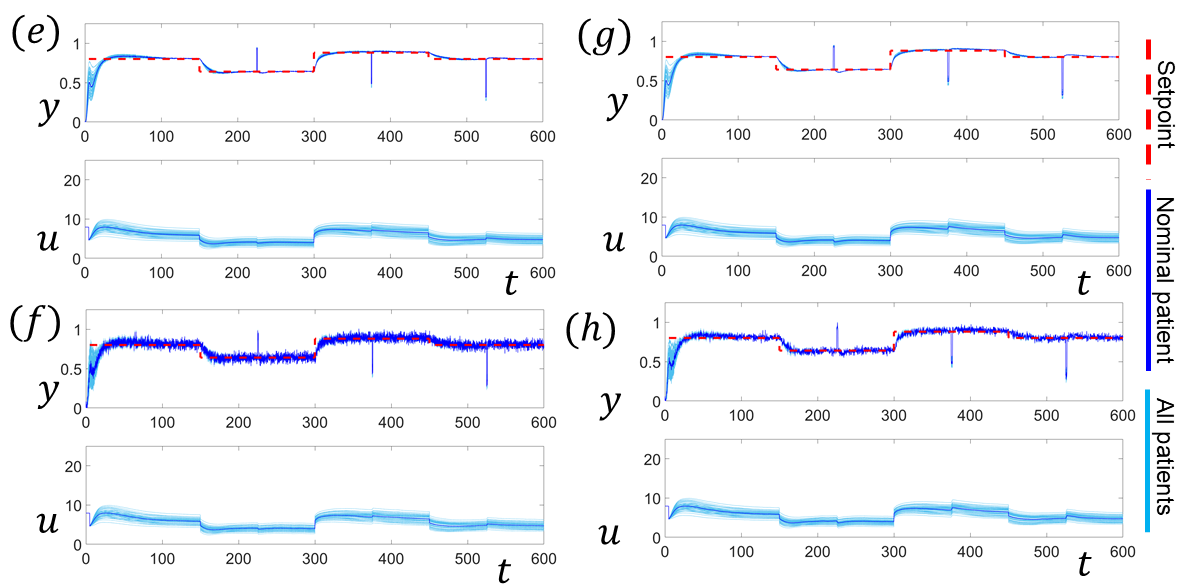}
 \caption{    \label{fig:LQG_IA_simu} {\bf ILQG}: Simulated trajectories for 45 patients (including the nominal subject, highlighted in solid blue) with 4 sets of ($60T_u, V$): (a,e) ($5\,s, 10^{-6}$), (b,f) ($5\,s, 10^{-4}$), (c,g) ($10\,s, 10^{-6}$), (d,h) ($10\,s, 10^{-4}$). (a-d) $\rho=10^{13}, K_I = 10$, (e-h) $\rho=10^{6}, K_I = 1$. See Table~\ref{tab:Scenario_B}.}
\end{figure}

\vspace{-5pt}
  For each control strategy we simulate for every patient set-point tracking with $T_u \in \lbrace 5/60\, min, 10/60\, min\rbrace$, and $V=\lbrace 10^{-6}, 10^{-4}\rbrace$ under conditions described in the following subsection. For each combination of $(T_u,V)$, we simulate: a PID-based CLAD (Fig.~\ref{fig:PID_simu}(a-d)), an ILQG-based CLAD with two control designs [$\rho = 10^{-13}, K_{I}=10$] (Fig.~\ref{fig:LQG_IA_simu}(a-d)) and [$\rho = 10^{-6}, K_{I}=1$] (Fig.~\ref{fig:LQG_IA_simu}(e-h)), and an LQG-based CLAD simulation under same conditions, but with $K_I=0$ (Fig.~\ref{fig:LQG_simu}(a-d) and Fig.~\ref{fig:LQG_simu}(e-h)). For each of these operating conditions, we simulate CLAD sessions for $N_p$ subjects such that the nominal plant and true plant are different. In these simulations, every subject is characterized by a respective $\Theta_P$ value, whereas controller parameters are characterized by a fixed set of parameters $\Theta_C$ determined based on a nominal model of the plant, $\Theta_{Pnom}$. To calculate the $\Theta_{Pnom}$, we consider a nominal subject (age $=35.5$ years, height $=177.5$ cm, weight $=79.5$ kg, male) using the median of each of these features.  The parameters, for this nominal patient are: $k_{10}=0.4282\,min^{-1}$, $k_{12}=0.4005\,min^{-1}$, $k_{13}=0.1958\,min^{-1}$, $k_{21}=0.0664\,min^{-1}$, $k_{31}=0.0035\,min^{-1}$, $k_{41}=0.4560\,min^{-1}$, $x_{50}=6.8\times 10^{-5}\,mg$, $\gamma = 3.0965$. All simulations are performed using \cite{MATLAB:2019b}/{Simulink}. To assess the performance across controller designs and across multiple simulation conditions, we use the following inaccuracy and bias metrics (\cite{dumont2009robust}),
\begin{align}
    \text{Inaccuracy, } & \epsilon_{I} = median\lbrace  100\vert (y_k - y_{sp})/y_{sp}\vert \rbrace
    \\
    \text{Bias, }& \epsilon_{B} = median\lbrace  100 (y_k - y_{sp})/y_{sp}  \rbrace
\end{align}
where, $y_k$ denotes the observation at the $k$-th time-point. We also report the average infusion, $u_{avg}$, calculated across each simulated session. 
\vspace{-5pt}
\subsection{Discussion}
\vspace{-5pt}
Our final assessment of controller performance across the simulated CLAD sessions is based on both the numerical outputs of $\epsilon_I$, $\epsilon_{B}$, $u_{avg}$ (Table~\ref{tab:Scenario_B}) and visual inspection of output and control trajectory plots (Figs.~\ref{fig:PID_simu}, \ref{fig:LQG_simu}, \ref{fig:LQG_IA_simu}). The inaccuracy and bias metrics for the PID are found to be lower than both ILQG and LQG.  
 The ILQG simulations show better set-point tracking performance relative to LQG, e.g. compare Fig.\ref{fig:LQG_simu}(a-d) vs. Fig. \ref{fig:LQG_IA_simu}(a-d) and Fig.\ref{fig:LQG_simu}(e-h) vs. Fig. \ref{fig:LQG_IA_simu}(e-h). Since the model mis-specification (difference between true and nominal models) has a significantly adverse effect on the performance metrics for Fig.~\ref{fig:LQG_simu}(e-h) this particular LQG design is unacceptable for any practical CLAD implementation. Also, from visual inspection we can infer that PID control leads to faster convergence to set-point relative to LQG or ILQG, but all three control strategies are able to provide stable regulation. The plots in Figs.~\ref{fig:PID_simu}(a-d),  \ref{fig:LQG_IA_simu}(a-d), and \ref{fig:LQG_IA_simu}(e-f) indicate that by lowering $\rho$ (the relative weighting of state cost to input cost in the LQG quadratic objective function) we can tune the ILQG to achieve a control signal that is not as sensitive to the observation noise and leads to tracking performance comparable to PID. In all the simulations except for Fig.~\ref{fig:LQG_simu}(e-h), the median $u_{avg}$ is found to be almost similar (at least to the first significant digit). A plausible reason for this observation is that in all cases, except Fig.~\ref{fig:LQG_simu}(e-h), the control strategies are able to converge on the same average infusion rate that would achieve and track the  same $y_{sp}$ for most part of the simulated sessions (as indicated by the median $\epsilon_I\le 4.2\%$ across all these sessions, which is markedly lower than the same metric in Fig.~\ref{fig:LQG_simu}(e-h)). Furthermore, the influence of the transient dynamics of the control signal due to set-point changes and noise become less pronounced in the total infused volume estimate (used to calculate $u_{avg}$) over such long sessions.  
 In the summary our aforementioned observations reflect the salient differences among the three control strategies. The PID controller is designed to have fast, stable dynamics on the output feedback error. On the other hand, the LQG controller is designed to minimize the quadratic cost Eq.~\eqref{eq:LQG_Obj} and a higher penalty on the deviation of state trajectory from $x_{sp}$ (relative to the deviation of control from $u_{sp}$) can lead to better set-point tracking of $y_{sp}$ by $y$. For ILQG, the addition of integral action leads to a sub-optimal performance relative to the LQG cost function but with improved steady state set-point tracking relative to LQG, and comparable to PID.
\vspace{-5pt}
\section{Conclusion}\label{sec:conclusion}
\vspace{-5pt}
In this work, we compared three control strategies: PID, LQG, and ILQG for a CLAD application. For the CLAD problem of set-point tracking of output from a single-input single-output system described by Eqs.~\eqref{eq:LinearPK} and \eqref{eq:sigmoidEmaxPD}, we draw the following inference from our analyses. PID can help with faster convergence to set-point when a set-point change occurs. Once a set-point is achieved, PID, ILQG and LQG can each provide stable set-point tracking. LQG requires a very high penalty on state cost relative to actuation cost to provide acceptable set-point tracking performance, but adding an integral action via an ILQG can lead to acceptable tracking performance with lower relative cost. In the presence of observation noise, ILQG can be tuned to achieve temporally smoother actuation with comparable set-point tracking performance relative to a PID controller. Our results indicate that an integral action, incorporated within a linear feedback control strategy, can improve set-point tracking of pharmacodynamic response in such a CLAD setting.

We envisage that this work will be useful to guide controller choices in future CLAD designs, and to numerically assess their performance prior to their usage in animal models or humans. Since the framework involving PK-PD model considered in this study is quite general (\cite{meibohm1997basic}), our work can be relevant to CLAD applications beyond the specific anesthetic drug and physiologic signal definitions of the assumed PK-PD model (\cite{schnider1998influence, schnider1999influence}). Furthermore, this paper may also serve as a tutorial introduction to applications of control theory in CLAD systems. 
\begin{ack}
\vspace{-5pt}
\label{sec:ackn}
We thank anonymous reviewers for helpful feedback and members of the Neuroscience Statistics Research Laboratory at MIT for helpful discussions.
\vspace{-5pt}
 \end{ack}

\small
\bibliography{ref.bib}       

\begin{thebibliography}{24}
\providecommand{\natexlab}[1]{#1}
\providecommand{\url}[1]{\texttt{#1}}
\providecommand{\urlprefix}{URL }
\expandafter\ifx\csname urlstyle\endcsname\relax
  \providecommand{\doi}[1]{doi:\discretionary{}{}{}#1}\else
  \providecommand{\doi}{doi:\discretionary{}{}{}\begingroup
  \urlstyle{rm}\Url}\fi

\bibitem[{Absalom et~al.(2011)Absalom, De~Keyser, and
  Struys}]{absalom2011closed}
Absalom, A.R., De~Keyser, R., and Struys, M.M. (2011).
\newblock Closed loop anesthesia: are we getting close to finding the holy
  grail?
\newblock \emph{Anesthesia \& Analgesia}, 112(3), 516--518.

\bibitem[{Absalom et~al.(2002)Absalom, Sutcliffe, and
  Kenny}]{absalom2002closed}
Absalom, A.R., Sutcliffe, N., and Kenny, G.N. (2002).
\newblock Closed-loop control of anesthesia using bispectral indexperformance
  assessment in patients undergoing major orthopedic surgery under combined
  general and regional anesthesia.
\newblock \emph{Anesthesiology: The Journal of the American Society of
  Anesthesiologists}, 96(1), 67--73.

\bibitem[{Astr{\"o}m and Murray(2010)}]{astrom2010feedback}
Astr{\"o}m, K.J. and Murray, R.M. (2010).
\newblock \emph{Feedback systems: an introduction for scientists and
  engineers}.
\newblock Princeton university press.

\bibitem[{Athans(1971)}]{athans1971role}
Athans, M. (1971).
\newblock The role and use of the stochastic linear-quadratic-gaussian problem
  in control system design.
\newblock \emph{IEEE transactions on automatic control}, 16(6), 529--552.

\bibitem[{Bickford(1950)}]{bickford1950automatic}
Bickford, R.G. (1950).
\newblock Automatic electroencephalographic control of general anesthesia.
\newblock \emph{Electroencephalography and Clinical Neurophysiology}, 2(1-4),
  93--96.

\bibitem[{Brown et~al.(2010)Brown, Lydic, and Schiff}]{brown2010general}
Brown, E.N., Lydic, R., and Schiff, N.D. (2010).
\newblock General anesthesia, sleep, and coma.
\newblock \emph{New England Journal of Medicine}, 363(27), 2638--2650.

\bibitem[{Crassidis and Junkins(2011)}]{crassidis2011optimal}
Crassidis, J.L. and Junkins, J.L. (2011).
\newblock \emph{Optimal estimation of dynamic systems}.
\newblock Chapman and Hall/CRC.

\bibitem[{Doyle(1978)}]{doyle1978guaranteed}
Doyle, J.C. (1978).
\newblock Guaranteed margins for lqg regulators.
\newblock \emph{IEEE Transactions on automatic Control}, 23(4), 756--757.

\bibitem[{Dumont and Ansermino(2013)}]{dumont2013closed}
Dumont, G.A. and Ansermino, J.M. (2013).
\newblock Closed-loop control of anesthesia: a primer for anesthesiologists.
\newblock \emph{Anesthesia \& Analgesia}, 117(5), 1130--1138.

\bibitem[{Dumont et~al.(2009)Dumont, Martinez, and
  Ansermino}]{dumont2009robust}
Dumont, G.A., Martinez, A., and Ansermino, J.M. (2009).
\newblock Robust control of depth of anesthesia.
\newblock \emph{International Journal of Adaptive Control and Signal
  Processing}, 23(5), 435--454.

\bibitem[{Grimble(1979)}]{grimble1979design}
Grimble, M. (1979).
\newblock Design of optimal stochastic regulating systems including integral
  action.
\newblock In \emph{Proceedings of the Institution of Electrical Engineers},
  volume 126, 841--848. IET.

\bibitem[{MATLAB(2019)}]{MATLAB:2019b}
MATLAB (2019).
\newblock \emph{Version: 9.7.0.1216025 (R2019b)}.
\newblock The MathWorks Inc., Natick, Massachusetts.

\bibitem[{Meibohm and Derendorf(1997)}]{meibohm1997basic}
Meibohm, B. and Derendorf, H. (1997).
\newblock Basic concepts of pharmacokinetic/pharmacodynamic (pk/pd) modelling.
\newblock \emph{International journal of clinical pharmacology and
  therapeutics}, 35(10), 401--413.

\bibitem[{Ngan~Kee et~al.(2007)Ngan~Kee, Tam, Khaw, Ng, Critchley, and
  Karmakar}]{ngan2007closed}
Ngan~Kee, W., Tam, Y., Khaw, K., Ng, F., Critchley, L., and Karmakar, M.
  (2007).
\newblock Closed-loop feedback computer-controlled infusion of phenylephrine
  for maintaining blood pressure during spinal anaesthesia for caesarean
  section: a preliminary descriptive study.
\newblock \emph{Anaesthesia}, 62(12), 1251--1256.

\bibitem[{Purdon et~al.(2013)Purdon, Pierce, Mukamel, Prerau, Walsh, Wong,
  Salazar-Gomez, Harrell, Sampson, Cimenser
  et~al.}]{purdon2013electroencephalogram}
Purdon, P.L., Pierce, E.T., Mukamel, E.A., Prerau, M.J., Walsh, J.L., Wong,
  K.F.K., Salazar-Gomez, A.F., Harrell, P.G., Sampson, A.L., Cimenser, A.,
  et~al. (2013).
\newblock Electroencephalogram signatures of loss and recovery of consciousness
  from propofol.
\newblock \emph{Proceedings of the National Academy of Sciences}, 110(12),
  E1142--E1151.

\bibitem[{Puri et~al.(2016)Puri, Mathew, Biswas, Dutta, Sood, Gombar, Palta,
  Tsering, Gautam, Jayant et~al.}]{puri2016multicenter}
Puri, G.D., Mathew, P.J., Biswas, I., Dutta, A., Sood, J., Gombar, S., Palta,
  S., Tsering, M., Gautam, P., Jayant, A., et~al. (2016).
\newblock A multicenter evaluation of a closed-loop anesthesia delivery system:
  a randomized controlled trial.
\newblock \emph{Anesthesia \& Analgesia}, 122(1), 106--114.

\bibitem[{Schnider et~al.(1998)Schnider, Minto, Gambus, Andresen, Goodale,
  Shafer, and Youngs}]{schnider1998influence}
Schnider, T.W., Minto, C.F., Gambus, P.L., Andresen, C., Goodale, D.B., Shafer,
  S.L., and Youngs, E.J. (1998).
\newblock The influence of method of administration and covariates on the
  pharmacokinetics of propofol in adult volunteers.
\newblock \emph{Anesthesiology: The Journal of the American Society of
  Anesthesiologists}, 88(5), 1170--1182.

\bibitem[{Schnider et~al.(1999)Schnider, Minto, Shafer, Gambus, Andresen,
  Goodale, and Youngs}]{schnider1999influence}
Schnider, T.W., Minto, C.F., Shafer, S.L., Gambus, P.L., Andresen, C., Goodale,
  D.B., and Youngs, E.J. (1999).
\newblock The influence of age on propofol pharmacodynamics.
\newblock \emph{Anesthesiology: The Journal of the American Society of
  Anesthesiologists}, 90(6), 1502--1516.

\bibitem[{Schwilden et~al.(1989)Schwilden, Stoeckel, and
  Sch{\"u}ttler}]{schwilden1989closed}
Schwilden, H., Stoeckel, H., and Sch{\"u}ttler, J. (1989).
\newblock Closed-loop feedback control of propofol anaesthesia by quantitative
  eeg analysis in humans.
\newblock \emph{BJA: British Journal of Anaesthesia}, 62(3), 290--296.

\bibitem[{Shafer and Gregg(1992)}]{shafer1992algorithms}
Shafer, S.L. and Gregg, K.M. (1992).
\newblock Algorithms to rapidly achieve and maintain stable drug concentrations
  at the site of drug effect with a computer-controlled infusion pump.
\newblock \emph{Journal of pharmacokinetics and biopharmaceutics}, 20(2),
  147--169.

\bibitem[{Shanechi et~al.(2013)Shanechi, Chemali, Liberman, Solt, and
  Brown}]{shanechi2013brain}
Shanechi, M.M., Chemali, J.J., Liberman, M., Solt, K., and Brown, E.N. (2013).
\newblock A brain-machine interface for control of medically-induced coma.
\newblock \emph{PLoS computational biology}, 9(10), e1003284.

\bibitem[{Weiser et~al.(2008)Weiser, Regenbogen, Thompson, Haynes, Lipsitz,
  Berry, and Gawande}]{weiser2008estimation}
Weiser, T.G., Regenbogen, S.E., Thompson, K.D., Haynes, A.B., Lipsitz, S.R.,
  Berry, W.R., and Gawande, A.A. (2008).
\newblock An estimation of the global volume of surgery: a modelling strategy
  based on available data.
\newblock \emph{The Lancet}, 372(9633), 139--144.

\bibitem[{Westover et~al.(2015)Westover, Kim, Ching, Purdon, and
  Brown}]{westover2015robust}
Westover, M.B., Kim, S.E., Ching, S., Purdon, P.L., and Brown, E.N. (2015).
\newblock Robust control of burst suppression for medical coma.
\newblock \emph{Journal of neural engineering}, 12(4), 046004.

\bibitem[{Yang et~al.(2019)Yang, Lee, Guidera, Vlasov, Pei, Brown, Solt, and
  Shanechi}]{yang2019developing}
Yang, Y., Lee, J.T., Guidera, J.A., Vlasov, K.Y., Pei, J., Brown, E.N., Solt,
  K., and Shanechi, M.M. (2019).
\newblock Developing a personalized closed-loop controller of medically-induced
  coma in a rodent model.
\newblock \emph{Journal of neural engineering}, 16(3), 036022.

\end{thebibliography}
\end{document}